\documentclass[twocolumn,showpacs,preprintnumbers,amsmath,amssymb,pre]{revtex4}
\usepackage{graphicx}
\usepackage{color}

\begin{document}

\title{Floating rings in vertical soap films : capillary driven bidimensional buoyancy  }

\author{N. Adami, H. Caps} 
\email{N.Adami@ulg.ac.be}
\affiliation{GRASP -- Universit\'{e} de Li\`{e}ge, physics departement B5, B4000 Li\`{e}ge, Belgium}

\date{\today}

\begin{abstract}
The present study aims to investigate the motion of buoyant rings in vertical soap films. Thickness differences and related bi-dimensional densities are  considered as the motor leading to bi-dimensional buoyancy. We show how this effect can be re-interpreted thanks to surface tension profiles in soap films. We propose a model involving surface tension profiles in order to describe the motion of buoyant particles in vertical soap films, and compare it to experimental data.    
\end{abstract}
\pacs{{\color{red}68.03.Cd, 68.15.+e, 47.15.gm}}


\maketitle

\section{Introduction}
Due to their particular fluid properties, soap films have been in the center of numerous studies these last decades \cite{Mysels, Couder1, Kellay1, Bruinsma, Nierstrasz, Scheid1, Wu2, Rivera1, Rutgers, Zhang_wu, Prasad}. Early works by Mysels et. al.  \cite{Mysels, Prasad, Schwartz} showed that their behaviours strongly depend on parameters such as surfactant solution viscosity \cite{Prasad}, but also on the nature of the surfactant used to make the films. Moreover, very specific surfactant-linked phenomena such as marginal regeneration \cite{Mysels, Nierstrasz} have also been spotted to account dramatically in phenomena such as drainage, leading to particularly large film lifetimes. All these interfacial properties have been widely used to model global behaviours of more complex system such as foams \cite{Weaire}. \\
In order to model natural and/or forced convection phenomena in vertical soap films (i.e. marginal regeneration), usual buoyancy must be reinterpreted to include interfacial phenomena. Experimental observation reveals that buoyant particles are usually thinner than their surroundings \cite{Couder1, Bruinsma}. These thickness differences are considered to be the motor of this nearly 2D buoyancy, and have been formalized using the 2D density defined as $\rho_2=\rho \Delta e $, with $\rho$ the density of the solution and $\Delta e$ the thickness difference. This interfacial/nearly-2D buoyancy have been successfully used to predict the results of various convection experiments in soap films \cite{Bruinsma, Zhang_wu, martin, Adami1}.\\
In this paper, we re-consider an experiments proposed by Couder and coworkers \cite{Couder1} in order to analyze the bi-dimensional buoyancy in vertical soap films. We show how mechanical equilibrium in soap film can be used to express bi-dimensional buoyancy in terms of surface tension profiles in soap films. A dynamical model is then proposed in order to predict the behavior of the rising rings as a function of the physical characteristics of both the rings and the soap films. 

\section{Experimental setup and materials}
\subsection{Soap films}
\begin{figure}[htbp]
\begin{center}
\includegraphics[height=5cm, width=7cm]{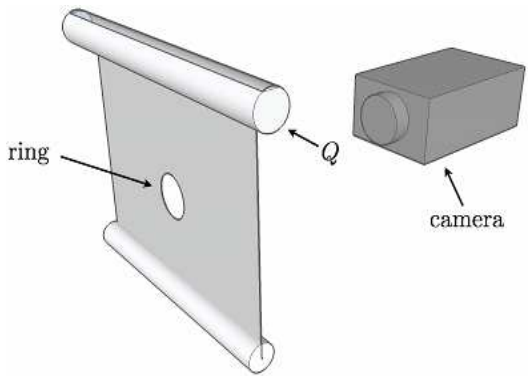} 
\caption{Experimental setup used to create maintained soap films.} 
\label{setup}
\end{center}
\end{figure}

Soap films are built from 3 \% of a solution of SLES+CAPB described in \cite{Denkov} plus 0.3 \% glycerol and double-distilled water. This mixture leads to typical density and viscosity of 1000 $\rm{kg/m^3}$ and $\rm{10^{-3}}$ Pa s, respectively. The surface tension $\gamma_0$ of the solution is determined to be $29.85 \pm 0.18$ mN/m. This mixture is used to generate a self-sustaining vertical soap film. In order to avoid the film thinning due to the gravitational drainage, we feed them thanks to a setup sketched on Fig.\ref{setup} \cite{Brunet1, Brunet2, Adami1}. A flow made of the soapy solution is injected with a constant flow rate $Q$ by both sides in a slit pipe, the slot pointing upward. When the soapy water gets out of the slot, it follows the edges of the pipe before reaching the top of the film which lies beneath the pipe. Doing so, it is possible to suppress the temporal evolutions of both thickness and surface tension profiles due to drainage, and to build soap films which can last for hours \cite{Adami1, Adami3}. Choosing the flow rate carefully allows to fix the thickness profile of the film. The thickness dependency versus the vertical coordinate $H$ has been determined using infrared absorption in \cite{Adami1}. It appeared that the thickness profiles of our films scale like a power law  $e(H)=aH^{-\beta}$ with $a$ being a constant and $\beta$ the scaling exponent. It can be shown that $\beta$ does not show any particular dependence in the flow rate $Q$ and remains close to -0.7 while $a$ increases with $Q$ from 1.8e-6 to 3e-6 $\mathrm{m^{\beta+1}}$, with $Q$ in $[1.6, 2.2]$ ml/min. Increasing $Q$ then leads to an increase of the thickness of the films without changing the shape of the profile. This effect has to be regarded as an intrinsic feature of the surfactant layers forming the film interfaces, and, further, of the surfactant itself. The corresponding surface tension profile can be expressed as (see \cite{Adami3} for details): 
\begin{equation}\label{gamma(H)}
\gamma(H)=\frac{\gamma_0}{2}+\frac{\rho g a}{2(1-\beta)}H^{1-\beta}
\end{equation}
\\
The lateral edges of the frame sustaining the film are made of stainless steel. All the films described in the present paper are $150\times 150$ $\rm{mm^2}$.\\ 

\subsection{Floating ring}
A simple way for evidencing the existence of surface tension is to introduce a soft ring into a soap film sustained by a frame \cite{Degennes}. At the beginning,  soap film are present both inside and outside the ring and the ring does not present any particular shape. But if the inner soap film is burst, the ring adopts a perfect circular shape. This can be explained by considering that a (surface) tension is present everywhere in the film. The inner film burst therefore leads to a radial force expressed as $2\gamma dl$ on every ring element $dl$, leading to the circular shape after the burst. This simple experiment attests from the existence of the surface tension linked to a soapy water interface.\\
Couder and coworkers \cite{Couder1} have shown that if the soap film is set verticaly and if the weight of the ring is weak enough, the burst of the inner film can lead to its upward motion into the vertical soap film. The burst of the inner film leads to the a thickness difference between the inside and outside of the ring, leading to the appearance of a bi-dimensional buoyant force on the ring. For the purpose of our experiments, we used human hairs which we glued on themselves so that they adopt a circular shape. In order to investigate the influence of the ring area, we built hairs rings of $8.69, 13.07, 14.56$ and $16.39$ mm in radius $R$. The hair diameter has been determined to be 100 $\rm{\mu m}$ using a high resolution microscope. Weighting the rings with a high accuracy  balance revealed that their masses are 5.1, 6.7, 8 and 9 $10^{-7}$ kg, respectively. Those measurements led to density and linear masse of $657$ kg/$\rm{m^3}$ and 1.23 $10^{-5}$ kg/m, respectively. Ring upward motions were imaged with a high speed camera, with a typical resolution of 1.5 $10^{-4}$ m and a acquisition frequency of 250Hz.\\

\section{Experimental results}
\begin{figure}[htbp]
\begin{center}
\includegraphics[width=9cm]{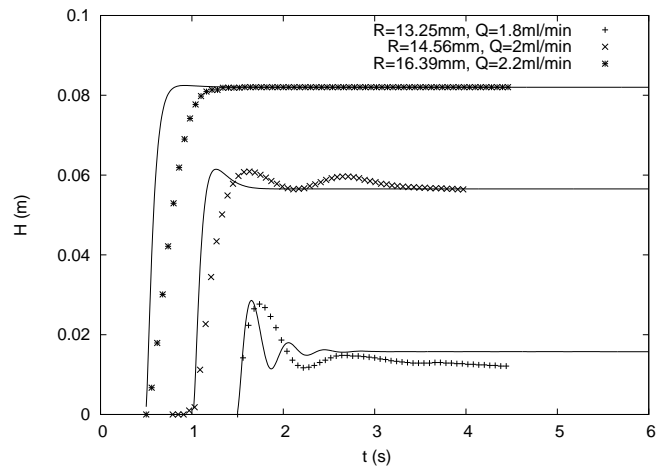}
\caption{Ring trajectories for different $(Q,R)$ couples. Points are experimental data and solid line are numerical trajectories using Eq. (\ref{eq_rise}). Time origin, considered as the burst time, has been shifted for more visibility.}
\label{exemple_rise}
\end{center}
\end{figure}

Filming the rises of the rings in the films allows to access to their position as a function of time for the different flows and radii used in our experiments. Rings are first introduced in the maintained soap film described in the previous section. The situation after introduction is so that soap films are present both inside and outside the ring, so that this latter is initially stuck against the bottom edge of the sustaining frame. We then burst the inner film and follow the resulting motion. Typical trajectories are show on Figure \ref{exemple_rise}, where time origin is defined as the moment when the inner film bursts. The ring position is defined from its center of mass. A fast uprise occurs  right after the burst, and is followed by a saturation of $H(t)$ to a equilibrium value called $H^*$. These curves clearly show that $H^*$ increases with both $Q$ and $R$. They also show that  the equilibrium linked to weak  and intermediate $(Q,R)$ values is preceded by damping oscillations of $H(t)$ around $H^*$, while higher  values lead to a critical saturation without any damping oscillations. \\

\section{Theory and modelling}
\subsection{Driving phenomenon}
\begin{figure}[htbp]
\begin{center}
\includegraphics[height=4cm, width=4cm]{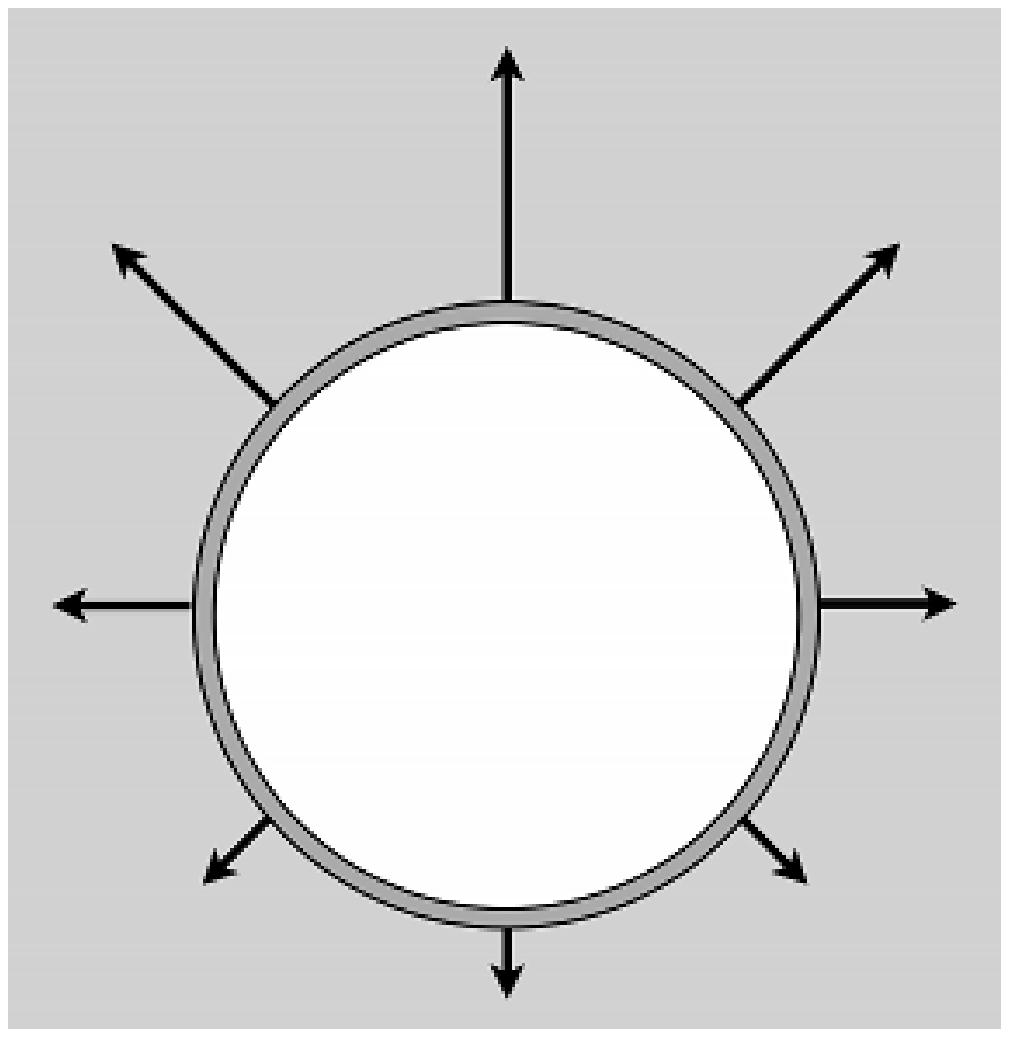}\ \includegraphics[width=3cm, height=4cm]{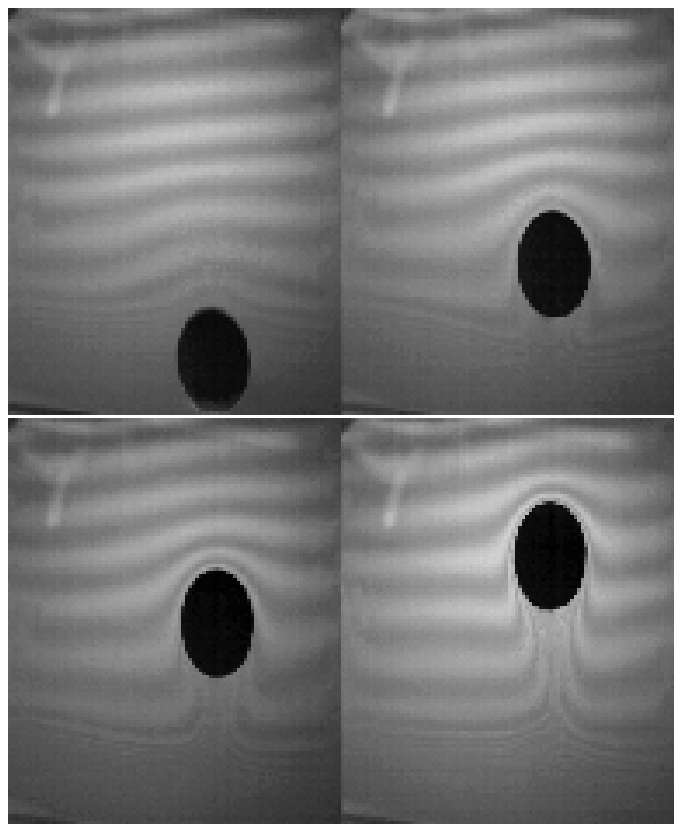}
\caption{Left : Schematic of surface tension forces acting on the ring (black arrows). Right : Shots of the ring during its rise. One can see the deformation of the fringe pattern of the soap film due to the rise of the ring.}
\label{surface_tension}
\end{center}
\end{figure}

\begin{figure}[htbp]
\begin{center}
\includegraphics[width=9cm]{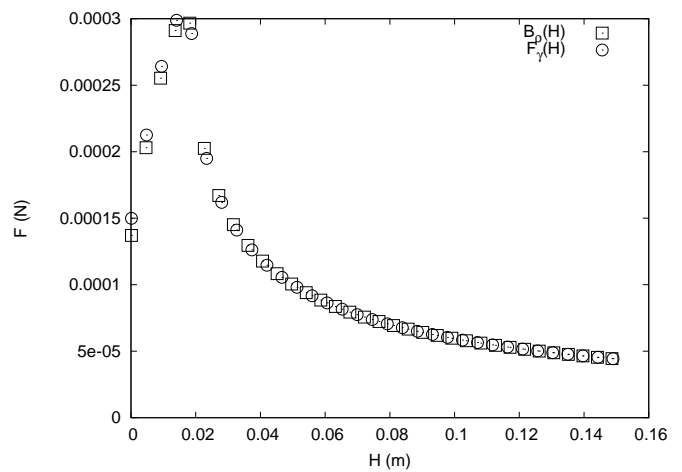}
\caption{Plot of $F_\gamma$ and $B_\rho$ versus $H$. Both forces have been computed with $R=22$ mm and $Q=2$ ml/min. Points have been shifted in a purpose of visibility.}
\label{match}
\end{center}
\end{figure}

In order to describe the driving force acting on the ring, Couder and coworkers proposed an adapted Archim\'ede-like expression of the usual buoyancy. Their proposition involves the thickness difference between the inner part of the ring and its surroundings, and reads : 

\begin{equation}\label{flott_e}
B_{\rho}=\Delta\rho Vg=\rho g\int_s e(H)ds
\end{equation} 
which is the usual expression of buoyancy where the usual $\Delta\rho$ is expressed as $\Delta\rho_2=\rho\Delta e$.  The variable $s$ is the area sustained by the ring and equals $\pi R^2$.  $B_\rho$ increases with $R$, which is consistent with experimental observations. Since thickness profiles are known \cite{Adami1}, $B_\rho$ can be computed as a function of $H$ for the different $R$ and $Q$ values. \\
Equation (\ref{flott_e}) expresses the buoyancy as being the weight of the displaced amount of liquid. Recent works have shown that vertical soap films interfaces must exhibit a vertical surface tension profile for mechanical equilibrium to be satisfied, i.e. the weight of the film must be balanced by surface tension forces everywhere in the soap film \cite{Adami3}. Applied to the ring with inner soap film present, or to any circular particle of the film, this mechanical equilibrium implies that the weight of the particle is carried by the surface tension forces provided by its surrounding. Figure \ref{surface_tension} illustrates the situation in the case of hair rings. Due to the surface tension gradient along the vertical coordinate, an upward resultant force acts on the ring. This force can be expressed as : 

\begin{equation}\label{force_gamma}
F_\gamma=2\oint \gamma\mathbf{u}\cdot\mathbf{n} dl
\end{equation}
with $\mathbf{n}$ the unitary normal vector defined by circular coordinate associated to the ring and $\mathbf{u}$ is a a unitary vector defined as $\mathbf{u}=-\mathbf{g}/g$. In terms of the weight-carrying nature of surface tension forces, this force must be equal to the weight of the film element defined by the surface $s$, which is actually given by Eq. (\ref{flott_e}). This mechanical consideration involves that buoyancy forces in soap films originate in surface tensions forces at the interface.  Since both thickness and related surface tension profiles of our films are known from experiments, we can compute Eq.(\ref{flott_e}) and Eq.(\ref{force_gamma}) versus $H$ and notice a matching of both curves, plotted on Figure \ref{match}. This matching confirms that 2D buoyancy is the result of a equilibrium between the mass of buoyant particles and surface tension forces linked to the film interfaces. To our knowledge, this is the first time that bi-dimensional buoyancy is interpreted as a surface tension-driven phenomenon.

\subsection{Floating ring dynamics}
Due to the saturating nature of Eqs. (\ref{flott_e}) and (\ref{force_gamma}) with $H$, it is normal for $H(t)$ to saturate as shown on Fig. \ref{exemple_rise}. Numerical computation of motion equation for the ring rise performed with $F_\gamma$ as driving force reveal to faithfully reproduce both typical rise time and $H^*$ involved in experiments. On the other hand, this simple dynamical description fails to reproduce the oscillations observed for weak and intermediate $(Q,R)$ couples. It is thus necessary to take additional forces into account to get a complete description of the dynamics of rising rings.  In addition to gravity, friction forces linked to the solution viscosity must be taken into account. In order to express them, we consider the basic definition of friction forces due to viscosity $F_f=\mu S dv/dy$, with $\mu$ the dynamic viscosity of the fluid, $S$ the surface on which friction forces act, $v$ the speed of the relative flow and $y$ the transverse coordinate along the direction of the flow. The typical velocity linked to the downward drainage, which can be estimated from flow conservation is the film as $Q/e(H)L$,  is 1cm/s. Experimental observation shows that the typical rising velocity of rings is of the order of 10 cm/s. The related Reynolds number is then of the order of 1000. In this regime, it is usual to consider that friction only acts on the top half of the ring. $S$ can then be expressed as $\pi RD$ as a consistent approximation. Since the rising velocity is one order of magnitude larger than the drainage velocity, it is consistent to approximate the velocity gradient $dv/dy$ as $v_{\rm{rel}}/D$, with $v_{\rm{rel}}=Q/e(H)L+dH/dt$. This leads to a general expression for the friction experienced by the ring, as : 

\begin{equation}\label{friction}
F_f=\mu \pi R\Big(\frac{Q}{e(H)L}+\frac{dH}{dt}\Big)
\end{equation}

As emphasized before, the typical diameter of hair used to build the rings is 100 $\rm{\mu m}$, whereas the thickness in the film is of the order of 10 $\rm{\mu m}$. Thus, a meniscus exists between the bulk of the film and the ring. As the ring begins to rise, the distance between the film interfaces must increase. The inverse phenomenon occurs on the lower half of the ring, where the distance between interfaces must decrease. A lubrication interaction must then be taken into account between the ring and the soap film. The typical pressure encountered by a flow of speed $U$ between plates separated by a distance $e$ is of the order of $\mu Ul/e^2$ with $l$ the length along which the flow occurs \cite{guyon}.  Considering that the lubrication interaction occurs on a distance equal to the air diameter, we can then express the lubrication force as : 

\begin{equation}\label{lub}
F_{lub}=\oint\frac{\mu D^2}{e^2(H)}\Big(\frac{Q}{e(H)L}+\frac{dH}{dt}\Big)\mathbf{u}\cdot \mathbf{n}dl 
\end{equation}

\noindent with previously defined notations.

\subsection{Mass of the rings}
\begin{figure}[htbp]
\begin{center}
\includegraphics[width=9cm]{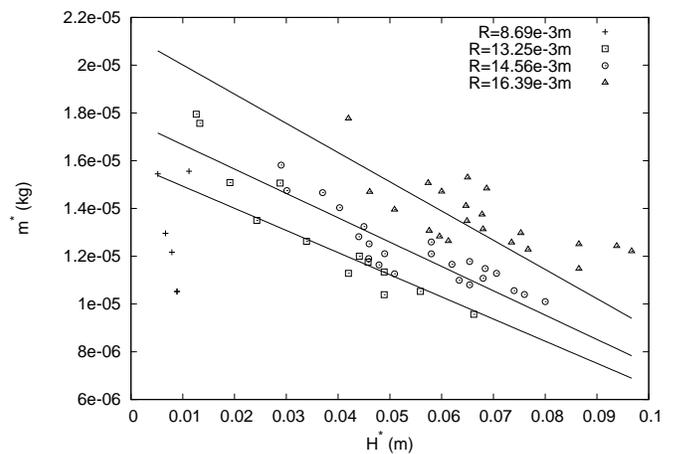}
\caption{Effective mass $m^*$ versus equilibrium height $H^*$. The larger the $(R,Q)$ couple, the larger the $m^*$. Dots are values obtained from mechanical equilibrium right after the rise of the ring, and solid lines are Eq.(\ref{mass}) plotted for the corresponding $(R,Q)$.}
\label{m*}
\end{center}
\end{figure}

After rise, and if the friction and lubrication forces are supposed to be negligible, the dynamical equilibrium for the ring writes $mg=F_\gamma(H^*)$. Introducing $H^*$ as obtained from experiments should enable to refine the mass of the rings. Computation of this equations bring masses which are two order of magnitude larger than the measured ring masses added from the mass of the meniscus they define with the soap film. Tests on numerical solving of our dynamical model revealed that relevant predictions can only be obtained by considering masses as determined from mechanical equilibrium after rise. Those values are plotted on Figure \ref{m*} as a function of $H^*$ for the different rings. To obtain a fully predictive model for ring rises, it is necessary to model the dependency of the effective masses $m^*$ as functions of $H$, $R$ and $Q$.\\  
Figure \ref{surface_tension} shows several shots of the ring during its rise. This image illustrates that the interference pattern in the film is disturbed by the rise of the ring, and that the deformation always reaches a given value of $H$ before the ring. These observations show that the motion of the ring in the film generates a fluid front, which seems to move with the ring. Momenta linked to upward and downward flows can be considered to evaluate the mass of this front. The meeting of both flows lead to a steady point localized at a distance $l$ from the ring. The fluid front thickness is assumed to be equal the $D$, so that its momentum can be expressed as $\rho\pi RDlv_r$, if $v_r$ is the typical rising velocity of the ring. This momentum opposes the momentum of the film column which lies above the ring, and which thus depends on its position in the film. The width of this column is $2R$. The downward momentum can thus be expressed as $$2R\rho\int_H^L e(H)v_d dh=2R\rho Q(L-H)/L$$ if the downward velocity $v_d$ is expressed as $Q/e(H)L$ thanks to mass conservation. Equaling both momenta allows to express the typical size of the fluid front as $l=2RQ(L-H)/L\pi RD\bar{v}$, with $\bar{v}$ being the mean velocity of the ring during its rise. This expression can be used to estimate the total mass of the rising ring as : 

\begin{equation}\label{mass}
m^*=m_0+\frac{2R \rho Q(L-H)}{L\bar{v}}
\end{equation}
This expression implies that the total mass of the ring must increase with both $Q$ and $R$, and that it must decrease with H. Figure \ref{m*} shows that Eq. (\ref{mass}) plotted for the different $(R,Q)$ couples is in good agreement with experimental values brought by dynamical equilibrium. 

\section{Comparison between model and experiments}
\begin{figure}[htbp]
\begin{center}
\includegraphics[width=9cm]{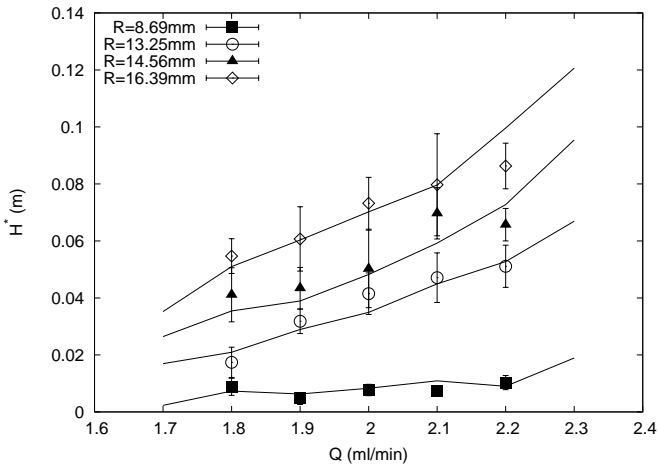}
\caption{Equilibriums height for the different rings versus injected flow. Dots are experimental data, while solid lines are prediction for $H^*$ obtained by solving Eq. (\ref{eq_rise}) thank to effectives masses $m^*$ predicted by Eq. (\ref{mass}). }
\label{H*}
\end{center}
\end{figure}

Since all the quantities involved in the dynamic equilibrium are known, we can then solve the Newtown equation for the rise of circular ring in vertical soap films. This equation reads : 

\begin{equation}\label{eq_rise}
m^*\frac{dH}{dt}=F_\gamma-F_f-F_{lub}-m^*g
\end{equation}
and  can be numerically solved thanks to a fourth-order Runge-Kutta method, leading to theoretical curves presented on Figure \ref{exemple_rise}. This figure shows that typical rise times, $H^*$ and damping oscillations for weak and intermediate $(Q,R)$ are faithfully reproduced by our model. Tests performed on the influence of the different forces showed that viscous friction is the agent responsible for damping oscillations, while lubrication forces tend to weakly decrease $H^*$ values. Those observation are in qualitative agreement with the directional nature of those forces.\\ Figure \ref{H*} illustrates the evolution of $H^*$ with $Q$ for different $R$ values. As the film becomes thicker with $Q$, surface tension gradients become stronger and a given ring rises further with increasing $Q$. For a fixed $Q$, larger $R$ lead to larger driving forces, and so to larger $H^*$. The solid lines on this Figure represent the predicted $H^*$ obtained from Eq.(\ref{eq_rise}), with $m^*$ obtained from Eq. (\ref{mass}), showing that experimental and theoretical data are in quantitative agreement, and that both modeling for the rising dynamics and effective masses $m^*$ are relevant in order to predict the behaviors of soft light rings set in entertained vertical soap films.  \\

The mechanical equilibrium of the ring right after the inner film burst can be used to estimate the minimal radius for the ring to rise in the film. Considering Eq.(\ref{eq_rise}) for t=0s (right after the burst of the inner film), the dynamic equilibrium for the ring writes $F_\gamma=mg+F_f+F_{lub}$. Those forces can be re-expressed for the particular case in t=0s as functions of $R$. In a purpose of simplicity, lubrication force are approximated by $F_{lub}\sim \mu U l/e^2 S$, with $S=2\pi RD$. After developments, the critical radius $R_c$ reads : 

\begin{equation}\label{radius}
R_c=\Big(\frac{ (\frac{\mu\pi Q}{eL}+2\pi\chi g+\rho g 2\pi D l_c+\frac{\mu Q l_c\pi D}{e^3L})(1-\beta)}{\rho g a I}\Big)^{\frac{1}{1-\beta}}
\end{equation}

with $\chi$ the linear mass of the air, $l_c$ the capillary length, and $I$ a constant value. The mass of the ring as been expressed as a sum of its mass outside the film plus the mass of the meniscus linking it to the film. Introducing typical values in this expression reveals that $R_c\sim 10^{-3}$ m, which is almost one order of magnitude smaller than the typical $R$ used in experiments.



\section{Conclusion}
We investigated the upward motion of light air rings in maintained vertical soap films. Experiments showed that the higher the radius of the ring and/or the flow used to maintain the film, the higher the equilibrium height. We showed that the usual buoyancy, expressed as the weight of the displaced amount of liquid, can be expressed as the resultant of surface tension forces acting on the ring, linked to the vertical surface tension profiles in the film, and that those forces are rigorously equivalent, as a consequence of the mechanical equilibrium considered in the film. We proposed a model, in which surface tension forces oppose gravity, viscous friction, and lubrication interaction, in order to describe the rise of the rings in the film. After numerical computation, we showed that this model brings faithful predictions about both rise dynamics and equilibrium height of the rings after growth. 

\textbf{Acknowledgements :} NA thanks FRS-FNRS for financial support. 





\begin{thebibliography}{99}
\bibitem{Adami3} Surface tension profiles in vertical soap films, To be published
\end{thebibliography}
\end{document}